# VANADIUM TRANSITIONS IN THE SPECTRUM OF ARCTURUS

M. P. Wood[1], C. Sneden[2], J. E. Lawler[3], E. A. Den Hartog[3], J. J. Cowan[4], and G. Nave[5]

[1] Department of Physics, University of St. Thomas, 2115 Summit Ave, St. Paul, Minnesota 55104; mpwood@stthomas.edu

[2] Department of Astronomy and McDonald Observatory, University of Texas, Austin, TX 78712; chris@verdi.as.utexas.edu

[3] Department of Physics, University of Wisconsin-Madison, 1150 University Ave, Madison, WI 53706; jelawler@wisc.edu; eadenhar@wisc.edu

[4] Homer L. Dodge Department of Physics and Astronomy, University of Oklahoma, Norman, OK 73019; jjcowan1@ou.edu

[5] National Institute of Standards and Technology, 100 Bureau Dr, Gaithersburg, MD 20899; gnave@nist.gov

ABSTRACT

We derive a new abundance for vanadium in the bright, mildly metal-poor red giant Arcturus. This star has an excellent high-resolution spectral atlas and well-understood atmospheric parameters, and it displays a rich set of neutral vanadium lines that are available for abundance extraction. We employ a newly recorded set of laboratory FTS spectra to investigate any potential discrepancies in previously reported V I log(*gf*) values near 900 nm. These new spectra support our earlier laboratory transition data and the calibration method utilized in that study. We then perform a synthetic spectrum analysis of weak V I features in Arcturus, deriving log $\varepsilon$(V) = 3.54 ± 0.01 ($\sigma$ = 0.04) from 55 lines. There are no significant abundance trends with wavelength, line strength, or lower excitation energy.

1. INTRODUCTION

A fundamental astrophysical goal is to derive detailed chemical compositions of Milky Way populations. What are these compositions as functions of time and spatial distribution in the Galaxy? How many different variables are needed to define major populations, and how many heterogeneous sub-groups really exist in each of these populations? Answers to these questions depend heavily on determining accurate elemental abundances for as many stars as possible.

In the past few decades there have been substantial improvements in both the quantity and quality of stellar abundances, and these will only improve in the future through the accumulation of data from many large-sample surveys (e.g., APOGEE, Majewski et al. 2015; Gaia-ESO, Gilmore et al. 2012; GALAH, Freeman et al. 2012). Better spectrographs, larger telescopes, more detailed model stellar atmospheres and line formation physics, and more robust atomic and molecular data have all contributed to substantial improvements in abundance accuracy. More recently, the development of HgCdTe (HAWAII) detector arrays has opened new infrared (IR) wavelength regions for astrophysics (e.g., APOGEE, Holtzman et al. 2015; IGRINS, Yuk et al. 2010). These regions are of particular importance in stellar astrophysics, where IR lines are generally better isolated and the continuum level better understood and easier to determine. This blossoming of IR astronomy puts pressure on the laboratory astrophysics community to keep pace.

Our group has concentrated on producing laboratory atomic data for as many lines as possible of more than 25 species of neutron-capture elements (atomic number $Z > 30$) and iron-group elements ($21 \leq Z \leq 30$). In our papers we have applied these new lab data (transition probabilities, hyperfine and isotopic substructures) to derive improved abundances of the solar photosphere and selected very metal-poor stars, those with [Fe/H][1] $< -2$. The choices of these stellar targets are well motivated astrophysically but they do not exploit the new lab data fully. In stellar spectra the number and strength of atomic absorption lines increases toward shorter wavelengths, and for those targets most abundance information is obtained at wavelengths shortward of Hα (656.3 nm). For example, our Sun has mostly weak atomic lines in the red and IR spectral regions, and many very metal-poor stars have almost no spectroscopic information in these spectral regions.

---

[1] We use standard abundance notations. For elements X and Y, the relative abundances are written $[X/Y] = \log_{10}(N_X/N_Y)_{star} - \log_{10}(N_X/N_Y)_{\odot}$. For element X, the "absolute" abundance is written $\log \varepsilon(X) = \log_{10}(N_X/N_H) + 12$. Metallicity is defined as [Fe/H]. The unit dex stands for decimal exponent such that x dex = $10^x$.

However, our lab data include many lines in the red and IR regions. To fully utilize these data we have begun to explore their application to the spectrum of the cool, mildly metal-poor red giant Arcturus, and in this paper we concentrate on neutral vanadium. Lawler et al. (2014; hereafter La14) reported new transition probabilities for 836 V I lines. Recently, Holmes et al. (2016; hereafter Ho16) have also determined transition probabilities for 208 lines of this species. For most lines in common between these two studies the results are in good agreement, but Ho16 suggest that the La14 log($gf$) values are systematically too small for six lines in common near 900 nm. Lines in this region are often used to construct calibration bridges between optical and IR spectra, so any issue with these lines needs to be resolved before laboratory IR work on V I beyond 1 μm can proceed confidently. As a result, this disagreement warrants further investigation.

In §2 we investigate the discrepancy noted by Ho16 using a newly recorded set of laboratory spectra to independently check our earlier calibration method, finding no cause for revision of the La14 data. In §3 the transition probabilities from La14 are applied to the spectrum of Arcturus. Finally, in §4-5 we discuss the new vanadium abundance in Arcturus in relation to previous work and suggest fruitful directions for future laboratory and stellar spectroscopy in this area.

## 2. LABORATORY BRANCHING FRACTION DATA REVISITED

Ho16 recently reported transition probabilities for 208 lines of V I, of which 188 are common to and in generally good agreement with our earlier work (La14). However, Ho16 noted a discrepancy for six lines in common around 900 nm, indicating that the La14 results for these transitions are too small by 0.18 dex on average and should be corrected. The noted discrepancy is limited to the $z^4D^o$ – $a^4P$ multiplet near 900 nm, and the relative strengths of the lines in the multiplet are not in dispute since these lines are in a nearly pure LS or Russell-Saunders multiplet. The multiplet branching fractions (BFs) are weak, BF < 0.04, compared to the dominant branches near 480 nm from the upper $z^4D^o$ levels. The total branching fraction for the multiplet is found to be approximately 0.032 in La14 and approximately 0.048 in Ho16.

This overall 0.016 BF discordance has very little effect on the log($gf$) values of other lines from the upper $z^4D^o$ levels.

To check the discordant multiplet, several new spectra (detailed in Table 1) of V/Ne and V/Ar hollow cathode (HC) lamps were recorded using the National Institute of Standards and Technology (NIST) 2-m Fourier transform spectrometer (FTS). Two of these spectra were recorded with sealed HC lamps from Heraeus[2] run at a current of 20 mA, while the rest result from a custom water-cooled high-current HC lamp. A tungsten standard lamp provided a relative radiometric calibration of these spectra, similar to the method employed by Ho16. One spectrum of a sealed V/Ar lamp (V032417.001) was recorded that provided independent confirmation of the calibration using Ar branching ratios as described below. However, the vanadium lines of interest in this spectrum were relatively weak and had uncertainties of 15 % to 20 %.

The log($gf$) agreement between the two earlier studies for shorter wavelength lines from the four $z^4D^o$ levels is very good, $<\log(gf)_{Ho16} - \log(gf)_{La14}> = 0.00 \pm 0.03$[3], and it is therefore reasonable to use branching ratios (BRs) to make a relative comparison between the multiplet lines in question and the other lines from these levels. We choose specifically to calculate the BR of lines from the $z^4D^o$ – a$^4$P multiplet around 900 nm compared to reference lines from the $z^4D^o$ – a$^4$D multiplet around 820 nm. BRs determined from spectra C through G in Table 1 are shown in Figure 1, where each panel represents a separate upper level and the different symbols represent BRs from Ho16, La14, and the present study. Error bars on the Ho16 results are derived using their quoted BF uncertainties. Error bars on the La14 results are derived in a similar way, with the 0.001 %/cm$^{-1}$ calibration uncertainty with respect to the dominant line (see La14 for details) removed from the BF uncertainties. The error bars on the BRs from this study include uncertainties in the integrated line intensities and in the relative radiometric calibration of the spectra. Each panel includes a single line for which all three studies are in perfect agreement at 1.0; this represents the reference line against which BRs are calculated for each upper level, and therefore this line does not include an error bar. These new spectra were analyzed using different software from that in La14 which, when combined with the different instrumentation and calibration method, results in an independent test of the La14 results. As expected, the BRs from La14, Ho16, and the present study are in good agreement for the lines near 820 nm. However, the Ho16 BRs for the 900 nm multiplet are systematically above the BRs from both La14 and this study, which are still in good agreement. This independent check of the La14 results suggests that the discordance for this multiplet is confined to the Ho16 results.

---

[2] Any commercial equipment, instruments, or materials are identified in this paper in order to specify the experimental procedure adequately. Such identification is not intended to imply recommendation or endorsement by NIST, nor is it intended to imply that the materials or equipment identified are necessarily the best available for the purpose.
[3] All uncertainties in this paper are one standard deviation.

Uncertainties in branching fraction studies covering a large wavelength range are often dominated by systematic uncertainties in the relative radiometric calibration. Ho16 used a tungsten standard lamp to determine their calibration, while the relative calibration of the La14 log(*gf*)s is based on Ar I and Ar II BRs from Whaling, Carle, & Pitt (1993) and earlier work by Hashiguchi & Hasikuni (1985), Danzmann & Kock (1982), and Adams & Whaling (1981). Ho16 claim the 0.18 dex discrepancy is the result of an error in this Ar BR calibration method, so it is worthwhile to compare the advantages and disadvantages of each method. The tungsten lamp is advantageous since it provides a continuum calibration. The response of the NIST 2-m FTS changes rapidly between 800 nm and 900 nm due to a dip in the reflectivity of aluminum in this region and Ar BRs may not have sufficient density to keep track of the changing response (see Figure 2). However, the tungsten lamp does not take into account the transmission of the window of the HC lamp. A separate measurement of the window transmission can be made if the lamp can be dismantled, but this is not possible with the sealed lamps used in the present study. The Ar BR method is advantageous in this case since it is recorded as part of the data spectrum and therefore the light is guaranteed to follow the same optical path, including through the HC lamp window. Additionally, any possible wavelength-dependent reflection/scattering of light in a HC is automatically captured in the Ar BR calibration (see Lawler & Den Hartog 2017). Hence the most reliable calibration can be obtained by combining the two methods, provided that the lines of interest can be excited in a lamp containing Ar.

Unfortunately no standard lamp spectra were recorded in conjunction with the V spectra from the National Solar Observatory digital archive analyzed by La14, but an independent check of the Ar BRs has recently been made. Figure 2 presents a comparison between a relative radiometric calibration based on 107 Ar I and Ar II lines (recorded simultaneously with a Ti HC lamp spectrum) and that of a tungsten standard lamp for spectrum A in Table 1. These earlier data were recorded using the NIST 2-m FTS as part of a separate study to investigate the primary tungsten standard for the FTS using the Ar BRs as a secondary calibration. Care was taken to ensure the FTS set-up was similar between the standard lamp spectrum and Ar spectrum, with the only changes being the orientation of the entrance mirror to select each lamp, the spectral resolution, and the gain of the FTS detectors. A change in the spectral resolution has no effect on the response of the spectrometer since it varies slowly with wavenumber. Changing the gain of the detectors can affect the response of the spectrometer if the amplifier gain settings have different bandwidths, but this is necessary due to the large difference in intensity between our standard lamp and the HC lamps. We have therefore measured the bandwidth of our detectors at different gain settings and corrected our response curve accordingly. In addition, two other corrections are incorporated in Figure 2 to ensure a fair comparison between the two calibrations: (1) the standard lamp includes a correction to account for the transmission of the HC window, and (2) the standard lamp includes an aging correction determined by comparison to a similar calibrated lamp with fewer hours of usage. There is generally good agreement between the Ar BR and tungsten standard lamp calibrations over the full calibrated range of the spectrum from 8500 cm$^{-1}$ to 24000 cm$^{-1}$. While there exist a few outliers, the large number of Ar calibration lines makes a comparison relatively straightforward. In particular, Figure 2 does not show a large systematic offset of the Ar BRs near 900 nm ($\approx$11000 cm$^{-1}$) and does not indicate the need for a correction to the Ar BR calibration as suggested by Ho16. A similar comparison of Ar BRs from spectrum B in Table 1 and shown in Figure 3 indicates that the calibration with the tungsten standard lamp agrees with that derived from the Ar BRs to within ±10 % in the region from 750 nm to 1100 nm (9000 cm$^{-1}$ to 13000 cm$^{-1}$). Ar BRs derived from this spectrum are compared to the literature values detailed above, and a value of 1.0 represent perfect agreement between derived and literature BRs. Due to the use of a sealed lamp with this spectrum, a window transmission correction could not be determined and applied to Figure 3, but the good agreement indicates that the change in transmission of the window across this region is small. The BRs of the vanadium lines from 800 nm to 900 nm in this spectrum also agree with La14 within their uncertainty of 15 % to 20 %.

Based on the independent verification of the La14 results and multiple checks of the Ar BR calibration method employed in that study, the oscillator strengths from La14 do not require any adjustment as suggested by Ho16.

## 3. THE VANADIUM ABUNDANCE OF ARCTURUS

Arcturus is a very bright red giant star ($V = -0.05$, $V-K = 2.96$) that has become an abundance standard for cool star spectroscopy. For example, Arcturus and the Sun are used as reference stars by the APOGEE team to test and improve infrared (H-Band) spectroscopic data (Shetrone et al. 2015). The optical spectrum of this star was studied extensively with the early photographic atlas of Griffin (1968). Arcturus now has an excellent public high-resolution, high signal-to-noise electronic spectrum[4] covering the ultraviolet (115 nm to 380 nm; Hinkle et al. 2005), optical (373 nm to 930 nm; Hinkle et al. 2000), and IR (0.9 μm to 5 μm; Hinkle & Wallace 2005) spectral regions. Hereafter, we refer collectively to these publications as the Arcturus Atlas. This spectrum is so line-rich that it is difficult to do effective abundance work at wavelengths below 500 nm. The line density is not as high in the yellow-red spectral region ($\lambda > 500$ nm), and inspection of the Arcturus Atlas reveals many strong and relatively unblended V I absorption features in this region.

La14 determined transition probabilities for 219 V I lines in the 500 nm to 920 nm wavelength domain. We searched the Arcturus Atlas for all of these lines, ignoring the line selection of La14 for their study of the Sun and the metal-poor star HD 84937. We first eliminated transitions that are severely blended and/or vanishingly weak. As in past papers of this series, to aid this effort we used the solar line compendium of Moore et al. (1966) and the comprehensive Kurucz (2011) atomic and molecular line database[5]. This process yielded about 100 V I lines that were judged worthy of further inspection. All of these remaining lines have $\lambda > 550$ nm since line blending by atomic and molecular ($C_2$, MgH) features block potentially useful lines blueward of this wavelength.

We generated synthetic spectra of small spectral regions around each of the remaining V I lines with the current version of the local thermodynamic equilibrium (LTE) plane-parallel line analysis code MOOG (Sneden 1973)[6]. Arcturus spectral analyses generally yield similar model parameters. In the PASTEL stellar parameter catalog (Soubiran et al. 2016)[7] simple means of the 42 entries for Arcturus yield effective temperature $<T_{eff}> = 4290 \pm 100$ K, surface gravity $<\log g> = 1.6 \pm 0.3$, and metallicity $<[Fe/H]> = -0.5 \pm 0.1$ (microturbulent velocities $v_t$ are not tabulated in this database). Recent detailed analyses include those of Ramírez & Allende Prieto (2011) who derived $T_{eff} = 4286$ K, $\log g = 1.66$, $v_t = 1.74$ km s$^{-1}$, and [Fe/H] = –0.52, and Peterson et al. (2017, and references therein), who recommend $T_{eff} = 4275$ K, $\log g = 1.3$, $v_t = 1.6$ km s$^{-1}$, and [Fe/H] = –0.55. We adopted the latter model for our abundance computations, and comment on this choice below. We interpolated a model atmosphere with the Peterson et al. (2017) parameters from the Kurucz (2011) model atmosphere grid[8] using software kindly provided by Andy McWilliam and Inese Ivans.

---

[4] Available at ftp://ftp.noao.edu/catalogs/arcturusatlas/
[5] http://kurucz.harvard.edu/linelists.html

Line lists for the synthetic spectra were generated as described in our previous papers (Sneden et al. 2016, and references therein). To summarize, we began with the atomic and molecular lines of the Kurucz (2011) database in 0.4 nm spectral intervals centered on the V I transition of interest. We then substituted or added to these lists the atomic transitions from our group's previous publications on iron-group and neutron-capture element species, CN from Sneden et al. (2014), C$_2$ from Ram et al. (2014), and MgH from Hinkle et al. (2013). As in La14 we expanded the V I lines to include hyperfine structure (HFS) components when these data were available. For this we searched the atomic physics literature anew, and in the Appendix we discuss the HFS data and provide a table that can be useful for synthetic spectrum computations.

Comparisons between these syntheses and the Arcturus Atlas resulted in the elimination of many candidate V I lines. Rejection of a line would result from one or more issues that made it unsuitable for abundance determinations: (1) the observed line was too weak (reduced equivalent width $\log(RW) = \log(EW/\lambda) < -6.2$); (2) the observed line was too blended with other atomic and/or molecular species (contaminant feature > 20 % of the total feature); (3) the synthesized line had a predicted strong blend in the Kurucz (2011) database that could not justifiably be neglected, even if it was not apparent in the observed line; (4) the Arcturus Atlas feature was too contaminated with very strong telluric $O_2$ or $H_2O$ features; or (5) the observed line was far too strong. The first four reasons for line rejection are standard for this kind of abundance work. Our justification for neglecting some of the strongest V I lines was to ensure that uncertainties in the assumed microturbulent velocity would not play a significant role in derived abundances, as discussed below.

---

[6] http://www.as.utexas.edu/~chris/moog.html
[7] http://pastel.obs.u-bordeauxatom.fr/
[8] http://kurucz.harvard.edu/grids.html

After eliminating these problematic lines, we subjected the remaining V I features to more detailed comparisons between synthetic and observed spectra. In Figure 4 we show examples of these comparisons. Each panel shows points from the Arcturus Atlas as well as syntheses assuming various V I abundances (the derived log ε abundance for the feature is noted in the panel). A red line shows the synthesis without any V I contribution, and a black line (often masked by the points) shows the best-fitting synthesis/observation match. The blue, green, and orange lines are for syntheses that assume V abundances that are factors of 1/4, 1/2, and 2 times the best-fit abundance, respectively (e.g., Δ log ε values that are –0.6, –0.3, and +0.3 shifted from the derived best-fit abundance for the feature). The three left-hand panels (a), (b), and (c) show representative lines in the yellow-red spectral region, and the three right-hand panels (d), (e), and (f) show lines near 900 nm for which Ho16 derive substantially different transition probabilities than La14. The gaps in the Arcturus Atlas points in panels (c) and (f) are small wavelength intervals where telluric absorptions are so deep that the underlying stellar spectrum cannot be reliably detected.

In our analysis we dealt only with relatively weak V I lines, ones with sensitivity to abundance but not microturbulent velocity $v_t$. We computed Arcturus V I curves-of-growth with our model atmosphere; these predict equivalent width (*EW*) as a function of assumed abundance for lines with no HFS substructure. The tests suggest that the very weak-line portion of the curve of growth extends up to residual width $\log(RW) = \log(EW/\lambda) \approx -5.5$. Residual widths show gradually decreasing sensitivity to abundance and increased dependence on assumed $v_t$ as line strength increases, and abundances for lines with $\log(RW) > -4.75$ are greatly dependent on the assumed microturbulent velocity. We measured approximate *EWs* for all potential transitions, and then conservatively chose to neglect V I lines stronger than $\log(RW) > -4.8$.

The wide HFS of many V I lines complicates statements about line strengths. HFS varies substantially from transition to transition, as illustrated in Figure 4. It is nearly negligible for the 554.5 nm line (panel a), but is about 0.05 nm for the 675.3 and 893.3 nm lines (panels b and d). For lines on the linear portion of the curve-of-growth, HFS broadens their profiles but does not change their total absorptions (and thus their *EW*s). However, HFS desaturates strong lines, increasing their total absorption while moving them toward the linear part of the curve of growth. Many of the stronger lines included have so much HFS broadening that they effectively act as unsaturated transitions. However, lines with $\log(RW) > -4.8$ systematically yield abundances about 0.05–0.15 dex lower than the weaker transitions. This trend can be cancelled by decreasing the microturbulent velocity to $v_t \approx 1.5$ km s$^{-1}$, but this would be a purely empirical correction. Exploration of this issue would involve a more thorough investigation of line formation in the Arcturus atmosphere and is beyond the scope of this paper.

In the end we employed 55 V I lines in the 550 nm to 920 nm spectral region to determine a new V abundance in Arcturus. Individual line abundances are listed in Table 2, derived from both the $\log(gf)$ values in La14 (columns 3 & 4) and in Ho16 (columns 5 & 6). From these lines we compute a mean value of $<\log \varepsilon> = 3.54 \pm 0.01$ (sample standard deviation $\sigma = 0.04$). La14 derived a solar photospheric abundance of $<\log \varepsilon_\odot> = 3.96 \pm 0.01$ ($\sigma = 0.04$), which yields [V/H] = –0.42, or [V/Fe] = +0.13. In Figure 5 we plot the individual line abundances as a function of wavelength. Red symbols are used to indicate the lines near 900 nm for which Ho16 derive $\log(gf)$ values that are on average 0.18 dex larger than those of La14. In this figure we write the abundance statistics separately for transitions below and above 890 nm. There is excellent agreement between the 49 lines at shorter wavelengths and the six lines near 900 nm. In short, there is no abundance trend with wavelength in Arcturus when the La14 transition probabilities are employed, as expected from the small $\sigma$ value of the mean abundance statistics. This agreement of derived V abundances near 900 nm and below this wavelength cannot be duplicated if the Ho16 $\log(gf)$ values are employed.

In Figure 6 we plot the line abundances as a function of transition strength. As in previous papers of this series we define relative strengths of lines of a single species as products of their transition probabilities and Boltzmann factors: $\log(gf) - \theta\chi$, where $\theta = 5040/T$. In our previous papers we have assumed $\theta = 1.0$ as a compromise between the solar value $\theta \approx 0.9$ and those of typical very metal-poor giants $\theta \approx 1.1$. Arcturus is much cooler, $\theta = 1.18$, and we have used this value for Figure 6. Inspection of Figure 6 indicates that there are no secular abundance trends with line strength, and the 900 nm lines span nearly ≈1 dex in transition strength. In Figure 7 we plot the V abundances as a function of lower level energy (excitation potential), and as with the previous figures there are no significant abundance trends. We note that all of the 900 nm lines arise from closely spaced lower energy levels near 1.2 eV, and these levels give rise to lines at shorter wavelengths that we have included in our abundance computations. Taken together, these figures show good agreement in derived V abundance for lines below λ < 826 nm and the lines near 900 nm which are questioned by Ho16. That there are no obvious trends with wavelength, transition strength, or excitation potential further supports our assertion that the data from La14 are consistent in both wavelength regions.

Ramírez & Allende Prieto (2011) reported V abundances in Arcturus by an EW analysis of eight weak V I lines, $\log(RW) = -6.0$ to $-5.3$, treating them as single absorbers. They reported [V/Fe] = +0.20 ± 0.05, which is consistent with our value given that they used different V I line parameters and a different model atmosphere than we employed. However, their preferred $v_t = 1.74$ km s$^{-1}$ leads to more abundance inconsistencies between weak and strong V I lines than we derived with the Peterson et al. value of $v_t = 1.6$ km s$^{-1}$. Further investigation of microturbulence in the Arcturus atmosphere is beyond the scope of our work.

## 4. NEUTRAL VANADIUM TRANSITIONS IN THE INFRARED

The longest-wavelength V I line reported by La14 is at 915.66 nm, but Hinkle et al. (1995) list 18 identifications of this species in their first Arcturus infrared atlas. Ho16 reports new log(*gf*) values for nine IR lines, but none of these appear in the Hinkle et al. list. We searched without success for these lines in the Arcturus Atlas. Nearly all of them lie in the thicket of telluric absorption that defines the division between photometric H and K bands. Only one of the Ho16 lines, 1798.75 nm, is not overly contaminated by neighboring telluric or stellar features. However, the V I absorption is less than 1 % deep at most in the observed Arcturus Atlas spectrum and thus can yield no reliable abundance value.

We explored this wavelength region with V I transitions listed in the Kurucz (2011) line compendium. We were able to detect many of the lines suggested by Hinkle et al. (1995), but the observed lines are weak. Having no lab data to guide this work, we adopted the Kurucz log(*gf*) values and made synthetic/observed spectrum matches of all clean V I lines that we identified. This exercise yielded V abundances in qualitative agreement with those determined from optical-region lines, but with large (±0.15) line-to-line scatter. This spectral region will be studied in the future with support from the new V I lab spectra acquired as part of this study.

## 5. VANADIUM TRENDS WITH METALLICITY

One of the important goals of our studies is to understand the changing nature of detailed stellar abundance ratios throughout the history of the Galaxy, i.e., Galactic Chemical Evolution (GCE). As detailed in §3, the present study indicates that the vanadium abundance in Arcturus is [V/H] = –0.42, or [V/Fe] = +0.13.

In Figure 8 we compare this newly determined abundance with other abundance surveys of Galactic stars including Feltzing & Gustafsson (1998), Barklem et al. (2005), Roederer et al. (2014), and Battistini & Bensby (2015); see also Reddy et al. (2006) for additional observations. These compilations include both halo and disk stars and extend over a more metal-rich region including super-solar, i.e., [Fe/H] > 0. At metallicities near [Fe/H] = –0.5 there is a small overabundance seen in [V/Fe], and our new value for Arcturus is consistent with that trend. Previous studies have also indicated modest [V/Fe] overabundances at lower metallicities typical of halo stars as illustrated in Figure 8 (see also Sneden et al. 2016).

GCE models have attempted to match these abundance observations as a function of metallicity. In Figure 8 we illustrate two such models, by Kobayashi et al. (2011) and a Kobayashi model with jet effects (Sneden et al. 2016). It is clear that these models mimic the general trends but are not yet a good fit to the observed V abundances. In particular it is seen that the deviation between GCE models and the abundance data grows larger at higher metallicities. We note, however, that (arbitrarily) adding 0.3 [V/Fe] to the predictions of the GCE model with jet effects would result in a good fit to almost all of the data, including the new abundance ratio for Arcturus.

Some other iron-group elements show a similar lack of fit with current GCE models (see Sneden et al. 2016). At metallicities typical of Arcturus the iron-group elements are produced by both core-collapse supernovae (SNe) and Type Ia SNe. Therefore, new measured values for iron-group abundances like that of V at these metallicities help to constrain the frequency and models of both SNe. Furthermore, additional high-resolution stellar observations spanning a wide metallicity range combined with abundances derived using precise laboratory transition data will provide meaningful inputs for improving GCE models and help our understanding of the time evolution of the elemental abundances in the Galaxy.

## 6. SUMMARY

The increasing availability of IR astrophysical spectra generates a significant need for accurate and reliable IR laboratory atomic transition data. A recent study of V I oscillator strengths by Ho16 found generally good agreement with an earlier study (La14). However, Ho16 noted disagreement with a small subset of lines in the near-IR around 900 nm, suggesting that the discordance indicated a calibration error by La14. In the present study, several new laboratory FTS spectra of V HC lamps were recorded at NIST to assess this discordance and compare the two calibration methods employed in the earlier studies. After analyzing these new data, we find no evidence for a problem with the La14 results or the Ar BR calibration technique near 900 nm. Further, we derived a new V abundance in the very bright red giant star Arcturus as a secondary test of the reliability of the La14 log(*gf*) values. That there is good abundance agreement among all V I lines utilized and that no obvious trends are seen in the abundances further bolsters our confidence in the earlier La14 results. Our new Arcturus [V/Fe] ratio is in good agreement with those of other mildly metal-poor stars. Finally, a thorough survey of the literature has resulted in a greatly expanded set of hyperfine structure component patterns for V I lines which may be of interest to future stellar abundance studies.

## ACKNOWLEDGMENTS

The authors thank the referee for their many helpful and detailed suggestions, which improved the quality of the paper. Our lab/stellar research has been supported by grants from NASA and NSF, most recently by NASA grant NNX16AE96G (J.E.L.), by NSF grant AST-1516182 (J.E.L., E.A.D.H.) and NSF grant AST-1211585 (C.S.). M.P.W. and J.E.L. served as guest researchers at NIST while recording the vanadium spectra analyzed in the present study. The earlier titanium spectrum was recorded while M.P.W. served as a NIST NRC Postdoctoral Fellow. J.J.C. acknowledges support by the National Science Foundation under Grant No. PHY-1430152 (JINA Center for the Evolution of the Elements).

APPENDIX

There has been substantial new work on hyperfine structure (HFS) of lines of V I since La14. Complete HFS line component patterns for a total of 94 lines of V I are provided in La14 based on both FTS measurements of lines connecting to 25 upper levels and published HFS data from Childs & Goodman (1967), Gough et al. (1985), Cochrane et al. (1998), Palmeri et al. (1995, 1997), Lefébvre et al. (2002), and Güzelçimen et al. (2011). The 94 lines selected in La14 were lines found to be useful in abundance determinations on the Sun and HD 84937. An article by Güzelçimen et al. (2014) appeared while La14 was in the Galley Proof stage that confirmed all but one of the FTS measurements in La14 to a fraction of an error bar. There is substantial variation in the accuracy and precision of HFS measurements depending on the technique used. The radio frequency technique as used by Childs and Goodman (1967), applicable to the ground and low metastable levels, yields results to better than 1 kHz. The Doppler-free laser technique as used by Gough et al. (1985) yields HFS constants to better than 1 MHz (30 MHz = 0.001 cm$^{-1}$). Nonlinear least-square fitting to FTS line profiles as used by La14 yields results accurate to around 0.001 cm$^{-1}$ to 0.002 cm$^{-1}$. For elemental abundance studies on the Sun and other stars, HFS component patterns to FTS accuracy are satisfactory. The additional publications employed herein to generate complete HFS line component patterns are Güzelçimen et al. (2014), Güzelçimen et al. (2015), Unkel et al. (1989), and Ho16 for the x$^6$D$_{3/2}$ level at 28368.753 cm$^{-1}$. The Casimir formula is from the text by Woodgate (1980)

$$\Delta E = \frac{AK}{2} + B\frac{3K(K+1) - 4I(I+1)J(J+1)}{8I(2I-1)J(2J-1)}$$

where $\Delta E$ is the shift in energy (or wavenumber) of a HFS sub-level (*F, J*) from the center of gravity of the fine structure level (*J*),

$$K = F(F+1) - J(J+1) - I(I+1),$$

*F* is the total atomic angular momentum, *J* is the total electronic angular momentum, and *I* = 3/2 is the nuclear spin for $^{51}$V. Although Childs & Goodman report HFS *C* (magnetic octupole) coefficients, their error bars overlap zero and they are not used here. Non-zero *B* (electric quadrupole) coefficients are included for some levels in this work, but only reliable HFS *A* (magnetic dipole) coefficients are necessary for abundance studies on stellar spectra.

Rather than add HFS component patterns for a small number of lines to the 94 published in La14, we decided to survey all 836 lines of V I with log($gf$)s in La14 and generate complete HFS component patterns for every line with a known upper and lower level HFS $A$ coefficient. In order to provide some added value in this exercise we compared the generated component patterns to our best FTS line profiles. Although nearly all of the hundreds of published HFS $A$ and $B$ coefficients were found to be reliable to FTS standards, several problems were identified and corrected in the published results. The HFS $A$ = 73 MHz coefficient of the $^2G_{9/2}$ level at 37361.95 cm$^{-1}$ reported by Güzelçimen et al. (2014) is not reliable and should be 0.0100 ± 0.001 cm$^{-1}$ (≈300 MHz) from our nonlinear least-square fits to FTS line profiles. Similarly, their HFS $A$ = 118 MHz coefficient of the $^4I_{11/2}$ level at 37315.93 cm$^{-1}$ is not reliable and should be 0.0103 (≈309 MHz) for astrophysical studies. This upper $^4I_{11/2}$ level lacks lines in our FTS data with resolved structure and we are not able to set a secure uncertainty on the new HFS $A$ coefficient. Finally, a line from the x$^2F_{3/2}$ level at 36766.041 cm$^{-1}$ is useful in Arcturus and has some partially resolved HFS in our FTS data. Nonlinear least-square fits to FTS line profiles yield 0.0100 ± 0.001 cm$^{-1}$ for the HFS $A$ of the x$^2F_{3/2}$ level, and this value is used with published HFS coefficients in the generation of line component patterns.

     Machine readable Table 3 has complete HFS line component patterns for more than 650 of the 839 lines with log($gf$)s in La14. Although only a few new HFS $A$ coefficients are used to generate the table, the complete patterns are convenient in astrophysical studies. This table represents a seven-fold expansion of the La14 table of HFS line component patterns and includes the majority of the lines in La14 with reliable log($gf$) values.

REFERENCES

Adams, D. L., & Whaling, W. 1981, JOSA, 71, 1036

Barklem, P. S., Christlieb, N., Beers, T. C., et al. 2005, A&A, 439, 129

Battistini, C., & Bensby, T. 2015, A&A, 577, 9

Childs, W. J., & Goodman, L. S. 1967, PhRv, 156, 64

Cochrane, E. C. A., Benton, D. M., Forest, D. H., & Griffith, J. A. R. 1998, JPhB, 31, 2203

Danzmann, K., & Kock, M. 1982, JOSA, 72, 1556

Eisenstein, D. J., Weinberg D. H. Agol E., et al. 2011. ApJ, 142, 72

Feltzing, S., & Gustafsson, B. 1998, A&AS, 129, 237

Freeman, K., Ness, M., Wylie-de Boer, E., et al. 2012, MNRAS, 428, 3660

Gilmore, G., Randich, S., Asplund, M., et al. 2012, Msngr, 147, 25

Gough, D. S., Hannaford, P., Lowe, R.M., & Willis, A. P. 1985, JPhB, 18, 3895

Güzelçimen, F., Başar, Gö, Öztürk, I. K., et al. 2011, JPhB, 44, 215001

Güzelçimen, F., Ypaici, B., Er, A., et al. 2014, ApJS, 214, 9

Güzelçimen, F., Er, A., Öztürk, I. K., et al. 2015, JPhB, 48, 115005

Griffin, R. F. 1968, A photometric atlas of the spectrum of Arcturus, [λ] 3600-8825 Å, Cambridge: Cambridge Philosophical Society

Hashiguchi, S., & Hasikuni, M. 1985, JPSJ, 54, 1290

Hinkle, K. H., & Wallace, L. 2005, Cosmic Abundances as Records of Stellar Evolution and Nucleosynthesis, ASP Conf. Ser., 336, 321

Hinkle, K. H., Wallace, L., & Livingston, W. 1995, Infrared Atlas of the Arcturus Spectrum 0.9 – 5.3 microns (San Francisco, ASP)

Hinkle, K., Wallace, L., Valenti, J., & Harmer, D. 2000, Visible and Near Infrared Atlas of the Arcturus Spectrum, 3727 – 9300 Å (San Francisco: ASP)

Hinkle, K., et al. 2005, Ultraviolet Atlas of the Arcturus Spectrum, 1150 – 3800 Å (San Francisco: ASP)

Hinkle, K. H., Wallace, L., Ram, R. S., et al. 2013, ApJS, 207, 26

Holmes, C. E., Pickering, J. C., Ruffoni M. P., et al. 2016, ApJS, 224, 35

Holtzman, J. A., Shetrone, M., Johnson, J. A., et al. 2015, AJ, 150, 148

Kobayashi, C., Karakas, A. I., & Umeda, H. 2011, MNRAS, 414, 3231

Kurucz, R. L. 2011, CaJPh, 89, 417

Lawler, J. E., & Den Hartog, E. A. 2017, JQSRT, submitted

Lawler, J. E., Wood, M. P., Den Hartog, E. A., et al. 2014, ApJS, 215, 20

Lefèbvre, P.-H., Garner, H.-P., & Biémont, E. 2002, PhyS, 66, 363

Majewski, S. R., Schiavon, R. P., Frinchaboy, et al. 2015, AJ, submitted, arXiv:1509.05420

Moore, C. E., Minnaert, M. G. J., & Houtgast, J. 1966, The Solar Spectrum 2935 Å to 8770 Å, (NBS Monograph Vol. 61; Washington, DC: US GPO)

Palmeri, P., Biémont, E., Aboussaïd, A., & Godefroid, M. 1995, JPhB, 28, 3741

Palmeri, P., Biémont, E., Quinet, P., et al. 1997, PhyS, 55, 586

Peck, E. R., & Reeder, K. 1972, JOSA, 62, 958

Peterson, R. C., Kurucz, R. L., & Ayres, T. R. 2017, ApJS, 229, 23

Ram, R. S., Brooke, J. S. A., Bernath, P. F., Sneden, C., & Lucatello, S. 2014, ApJS, 211, 5

Ramírez, I., & Allende Prieto, C. 2011, ApJ, 743, 135

Reddy, B. E., Lambert, D.L., & Allende Prieto, C. 2006, MNRAS 367, 1329

Roederer, I., Preston, G. W., Thompson, I. B., et al. 2014, AJ, 147, 136

Shetrone, M., Bizyaev, D., Lawler, J.E., Allende Prieto, C., Johnson, J. A., et al., ApJS 221, 24 (2015)

Sneden, C., Lucatello, S., Ram, R. S., Brooke, J. S. A., & Bernath, P. 2014, ApJS, 214, 26

Sneden, C., Cowan, J. J., Kobayashi, C., et al. 2016, ApJ, 817, 53

Thorne, A. P., Pickering, J. C., & Semeniuk, J. I. 2011, ApJS, 192, 11

Unkel, P., Buch P., Dembczynski, J., 1989, Z. Phys. D, 11, 259

Whaling, W., Carle, M. T., & Pitt, M. L. 1993, JQSRT, 50, 7

Woodgate, G. K. 1980, Elementary Atomic Structure (2nd ed.; Oxford: Clarendon Press), 184

Yuk, I.-S., Jaffe, D. T., Barnes, S., et al. 2010, in Proc. SPIE 7735, Ground-based and Airborne Instrumentation for Astronomy III, 77351M

FIGURE CAPTIONS

Figure 1. Branching ratios for lines from each of the upper $z^4D^o$ levels to the a$^4$D and a$^4$P lower levels. The orange squares, black crosses, and green circles are results from Ho16, La14, and the present study, respectively. Associated error bars are included for all three studies. The single line for which all three symbols perfectly overlap at 1.0 is the reference line against which all BRs for a given upper level (specified in the upper right corner) are calculated.

Figure 2. A relative radiometric calibration of the 2-m FTS at NIST based on 107 Ar I and Ar II lines (blue open circles) to a W strip standard lamp calibration (black line) in the near-IR and optical. The grey shaded region represents the uncertainty in the standard lamp calibration combined in quadrature with the HC window transmission uncertainty. Error bars represent individual Ar line intensity uncertainties combined in quadrature with the gain correction uncertainty.

Figure 3. Ratio of Ar BRs derived from NIST 2-m FTS spectra calibrated using a tungsten standard lamp to Ar BRs in the literature (see text for details). The different colors represent BRs from different upper levels in Ar I and Ar II. A ratio of 1.0 indicates that the BRs derived using the standard lamp calibrated spectra are in perfect agreement with the literature. The symbols and colors represent different line groups, with numbers taken from Whaling et al. (1993).

Figure 4. Representative Arcturus Atlas spectra (open circles) of V I transitions and synthetic spectra (lines). In each panel the black line represents the best fit synthesis to the observation, and its assumed V abundance is written in the lower right of the panel. The orange line is the synthesis with the V abundance increased by 0.3 dex, the green line for a decrease of 0.3 dex, the blue lines for a decrease of 0.6 dex, and the red line is for no V contribution to the feature.

Figure 5. Abundances from V I lines derived using La14 transition data, plotted as a function of wavelength. Abundances below and above 890 nm are color-coded, and their means and standard deviations are written in the figure. The solid horizontal line is drawn at the mean value for the lines with λ < 826 nm, and the dotted lines are drawn at ± 1 σ.

Figure 6. Abundances from V I lines derived using La14 transition data, plotted as a function of transition strength. Symbols and lines are as in Figure 5. See text for further explanation of the transition strength definition adopted here.

Figure 7. Abundances from V I lines derived using La14 transition data, plotted as a function of lower level energy (excitation potential). Symbols and lines are as in Figure 5.

Figure 8. The newly determined V abundance in Arcturus compared to abundance values of other Galactic disk and halo stars. (See text for details.)

Table 1. Newly recorded Fourier transform spectra of hollow cathode (HC) lamps recorded using the NIST 2-m FTS. All spectra are recorded using Si detectors, silver optics, and an aluminum beamsplitter, and have a useable wavenumber range of approximately 8500 cm$^{-1}$ to 24000 cm$^{-1}$.

| Index | Spectrum[a] | Serial Number | HC Lamp Type[b] | Buffer Gas | Lamp Current (mA) | Limit of Resolution (cm$^{-1}$) | Coadds |
|---|---|---|---|---|---|---|---|
| A | Ti041715b | 001 | Water-cooled | Ar | 500 | 0.015 | 25 |
| B | V032417 | 001 | Sealed | Ar | 20 | 0.08 | 64 |
| C | V062117a | 001 | Sealed | Ne | 20 | 0.025 | 32 |
| D | V062317a | 001 | Water-cooled | Ne | 200 | 0.03 | 100 |
| E | V062317a | 002 | Water-cooled | Ne | 700 | 0.025 | 115 |
| F | V062617b | 001 | Water-cooled | Ar | 250 | 0.02 | 95 |
| G | V062617b | 002 | Water-cooled | Ar | 500 | 0.02 | 60 |

[a] The date can be determined from the numbers in the spectrum name (MM/DD/YY).
[b] Lamp types include commercially-available sealed HC lamps from Heraeus and a high-current water-cooled HC lamp with a demountable window.

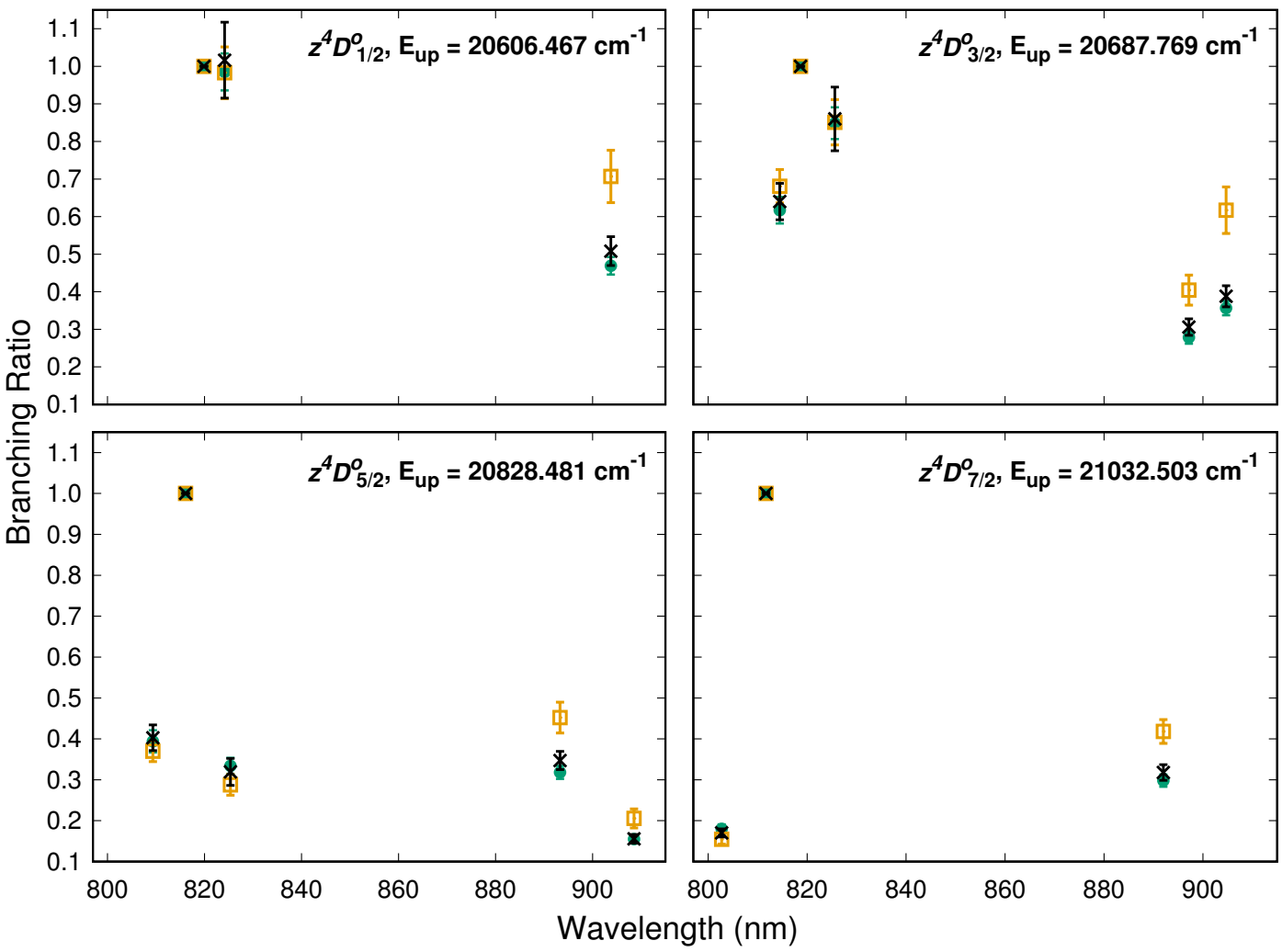

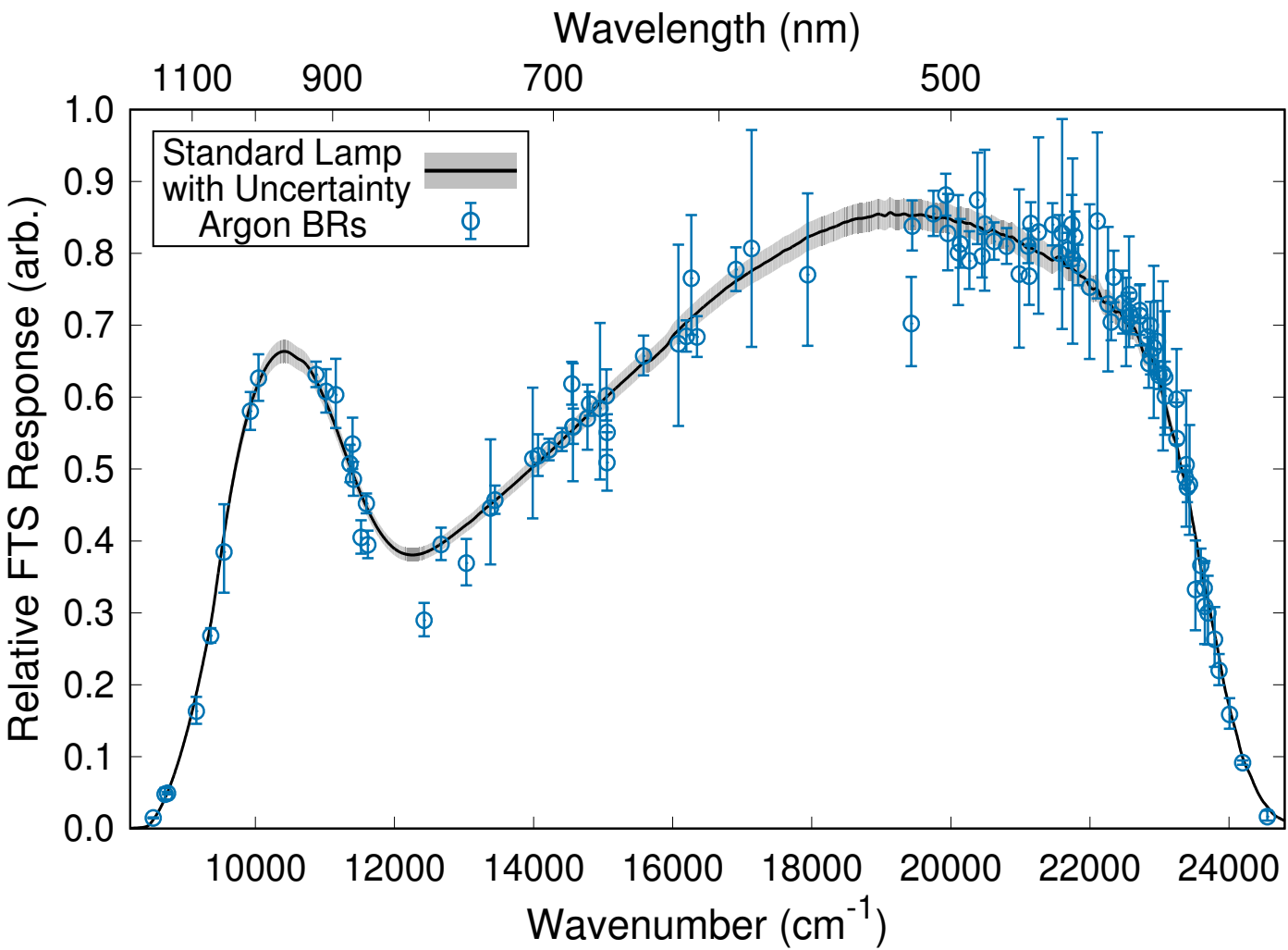

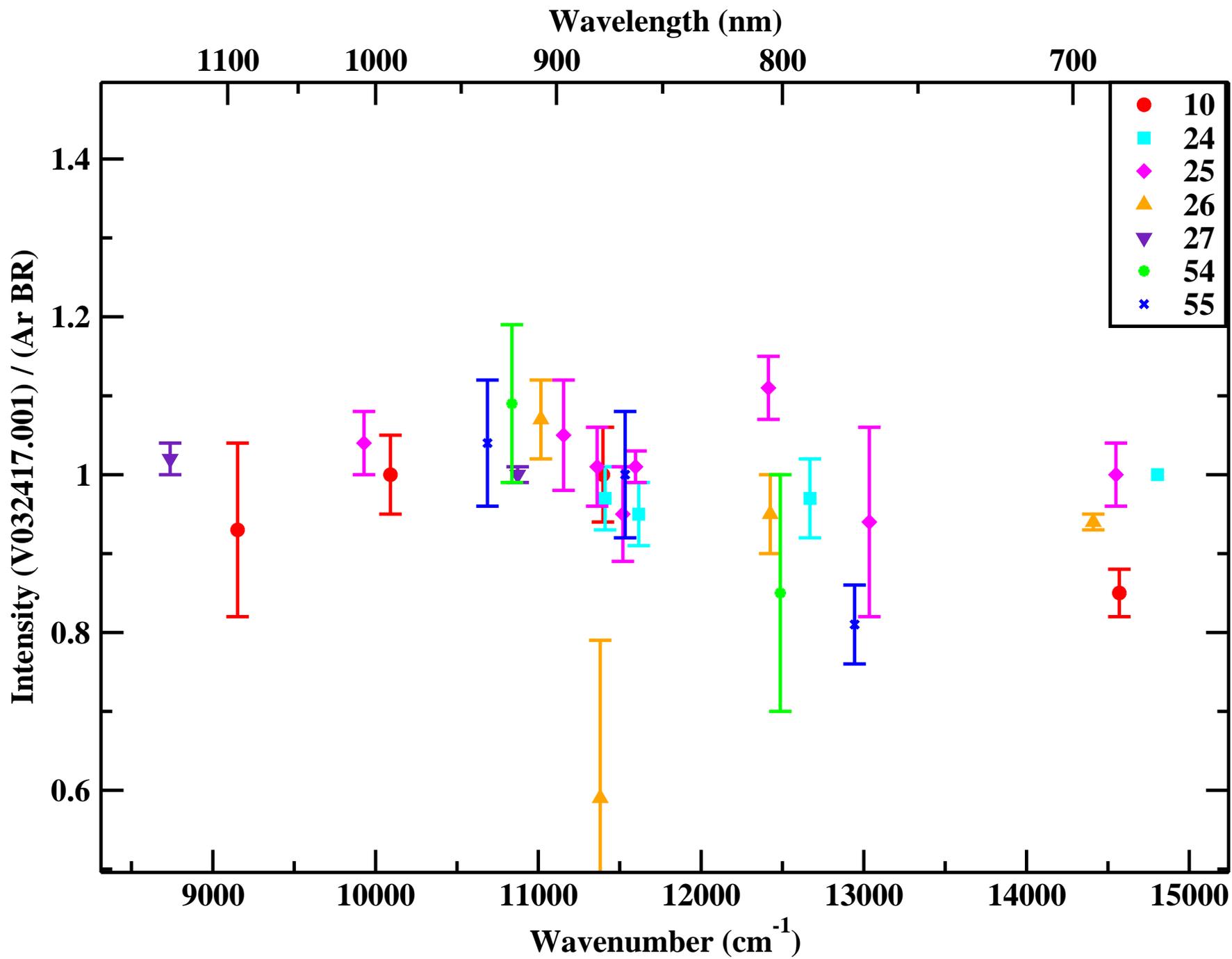

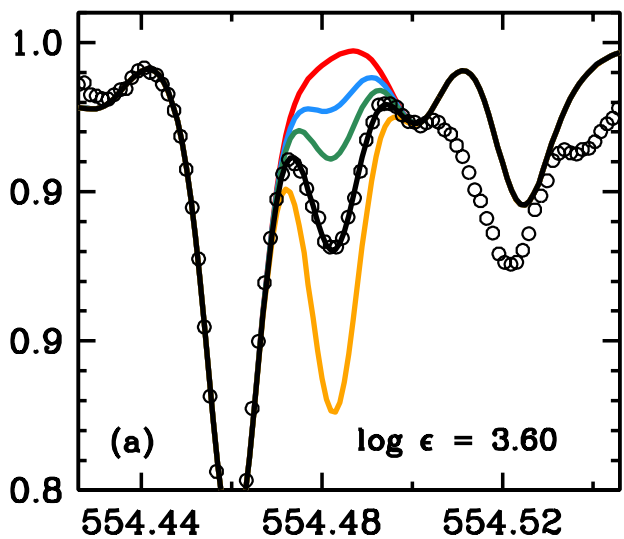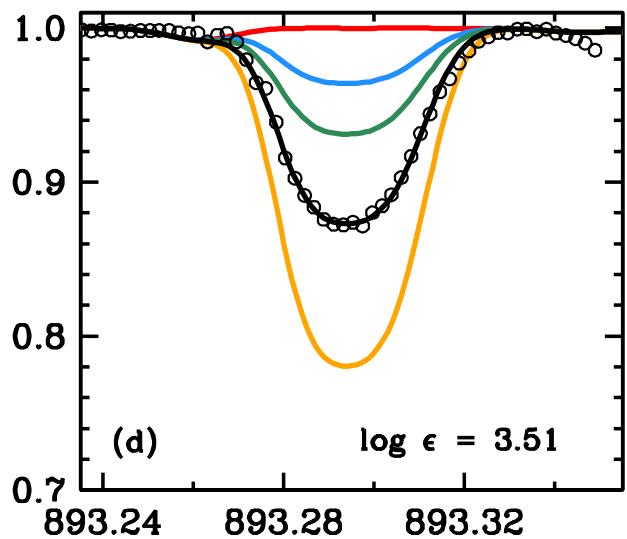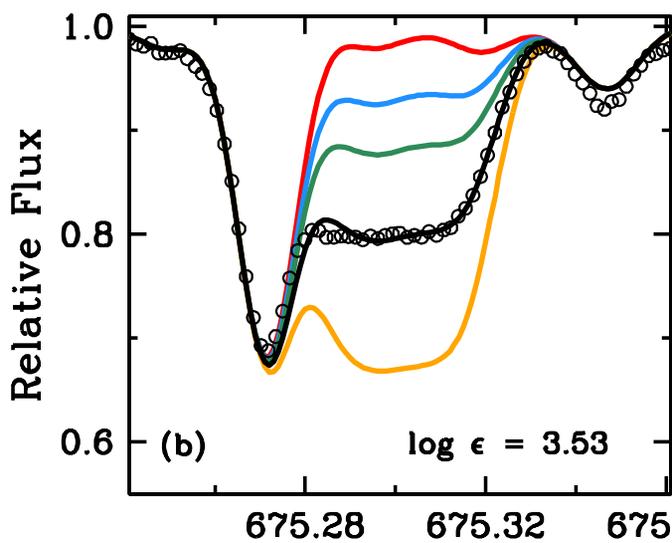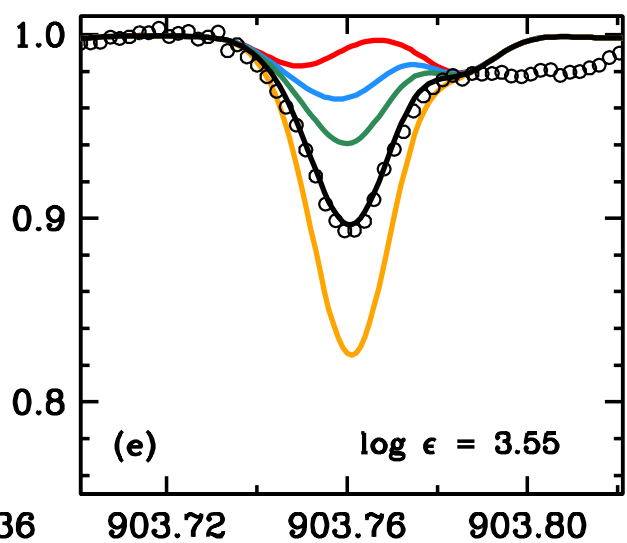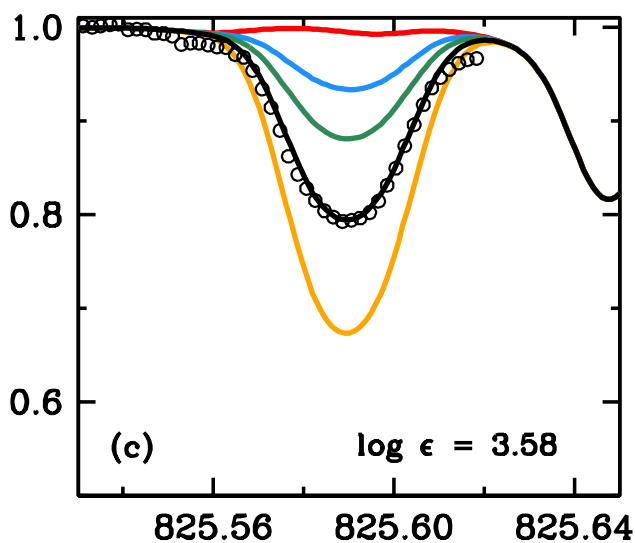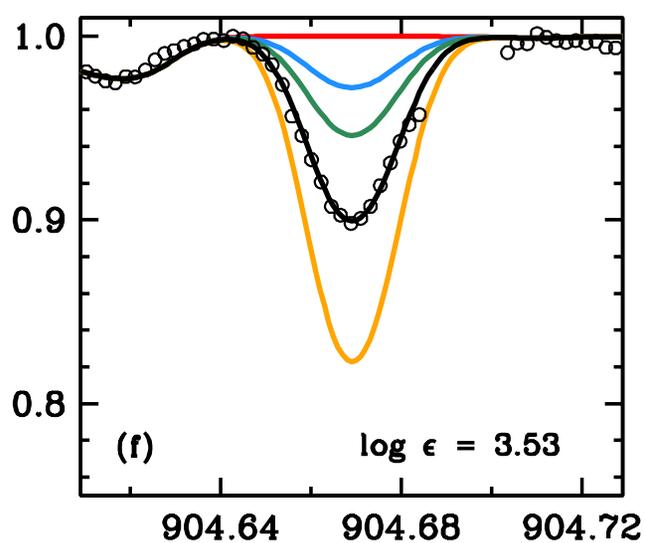

Table 2. Vanadium Abundances from V I lines in Arcturus.

| λ (nm) | E$_{lower}$ (eV) | log(gf) La14 | log ε La14 | log(gf) Ho16 | log ε Ho16 |
|---|---|---|---|---|---|
| 554.4858 | 1.051 | -2.56 | 3.60 | -2.57 | 3.61 |
| 555.7450 | 0.017 | -3.45 | 3.52 | -3.47 | 3.54 |
| 556.0547 | 0.040 | -3.63 | 3.53 | ... | ... |
| 557.6502 | 1.064 | -2.45 | 3.57 | -2.49 | 3.61 |
| 557.8373 | 1.051 | -2.58 | 3.55 | ... | ... |
| 558.4497 | 1.064 | -1.18 | 3.48 | ... | ... |
| 558.8457 | 1.081 | -2.25 | 3.55 | -2.30 | 3.60 |
| 559.2960 | 0.040 | -3.21 | 3.46 | -3.24 | 3.49 |
| 559.7797 | 1.064 | -2.42 | 3.61 | -2.57 | 3.76 |
| 560.4935 | 1.043 | -1.26 | 3.48 | -1.18 | 3.40 |
| 563.2454 | 0.069 | -3.23 | 3.51 | -3.29 | 3.57 |
| 568.7753 | 2.379 | -1.14 | 3.58 | ... | ... |
| 572.5639 | 2.365 | -0.17 | 3.55 | ... | ... |
| 575.0642 | 1.955 | -1.29 | 3.56 | ... | ... |
| 577.2412 | 1.931 | -0.55 | 3.53 | ... | ... |
| 577.6687 | 1.081 | -1.54 | 3.53 | ... | ... |
| 600.2303 | 1.218 | -1.79 | 3.51 | ... | ... |
| 601.7920 | 1.195 | -2.39 | 3.59 | -2.34 | 3.54 |
| 605.8142 | 1.043 | -1.40 | 3.48 | -1.37 | 3.45 |
| 608.7496 | 1.051 | -2.51 | 3.53 | ... | ... |
| 618.1862 | 1.051 | -2.45 | 3.54 | ... | ... |
| 619.0506 | 1.064 | -2.53 | 3.53 | ... | ... |
| 624.3107 | 0.301 | -0.94 | 3.48 | -0.95 | 3.49 |
| 628.5160 | 0.275 | -1.54 | 3.47 | -1.55 | 3.48 |
| 632.6836 | 1.868 | -0.76 | 3.53 | ... | ... |
| 634.9473 | 1.853 | -0.95 | 3.57 | ... | ... |
| 635.5572 | 2.122 | -1.42 | 3.58 | ... | ... |
| 635.7291 | 1.849 | -1.06 | 3.58 | ... | ... |
| 645.2344 | 1.195 | -1.22 | 3.53 | -1.18 | 3.49 |
| 653.1421 | 1.218 | -0.85 | 3.48 | -0.82 | 3.45 |
| 654.3504 | 1.195 | -1.71 | 3.51 | -1.66 | 3.46 |
| 656.5883 | 1.183 | -2.05 | 3.51 | -2.05 | 3.51 |
| 660.5974 | 1.195 | -1.34 | 3.51 | -1.31 | 3.48 |
| 660.7828 | 1.350 | -1.98 | 3.63 | ... | ... |
| 662.4845 | 1.218 | -1.30 | 3.48 | -1.26 | 3.44 |
| 675.3010 | 1.081 | -1.50 | 3.53 | ... | ... |
| 678.5000 | 1.051 | -1.88 | 3.55 | ... | ... |
| 681.2400 | 1.043 | -2.14 | 3.55 | ... | ... |
| 683.2442 | 1.064 | -2.43 | 3.53 | ... | ... |
| 684.1877 | 1.051 | -2.53 | 3.56 | ... | ... |
| 733.8931 | 2.138 | -0.74 | 3.56 | ... | ... |
| 736.3154 | 2.122 | -1.01 | 3.53 | ... | ... |
| 802.7366 | 1.064 | -1.85 | 3.55 | -1.86 | 3.56 |
| 809.3468 | 1.051 | -1.76 | 3.60 | -1.76 | 3.60 |
| 811.6789 | 1.081 | -1.07 | 3.57 | -1.03 | 3.53 |
| 814.4560 | 1.043 | -1.90 | 3.56 | -1.87 | 3.53 |
| 824.1599 | 1.051 | -1.90 | 3.58 | -1.87 | 3.55 |
| 825.3506 | 1.081 | -1.85 | 3.58 | -1.84 | 3.57 |
| 825.5896 | 1.064 | -1.75 | 3.58 | -1.70 | 3.53 |
| 891.9847 | 1.218 | -1.49 | 3.58 | -1.33 | 3.42 |
| 893.2947 | 1.195 | -1.74 | 3.51 | -1.58 | 3.35 |
| 897.1673 | 1.183 | -2.13 | 3.57 | -1.98 | 3.42 |
| 903.7613 | 1.183 | -2.12 | 3.55 | -1.94 | 3.37 |
| 904.6693 | 1.195 | -2.02 | 3.53 | -1.79 | 3.30 |
| 908.5231 | 1.218 | -2.08 | 3.55 | -1.91 | 3.38 |

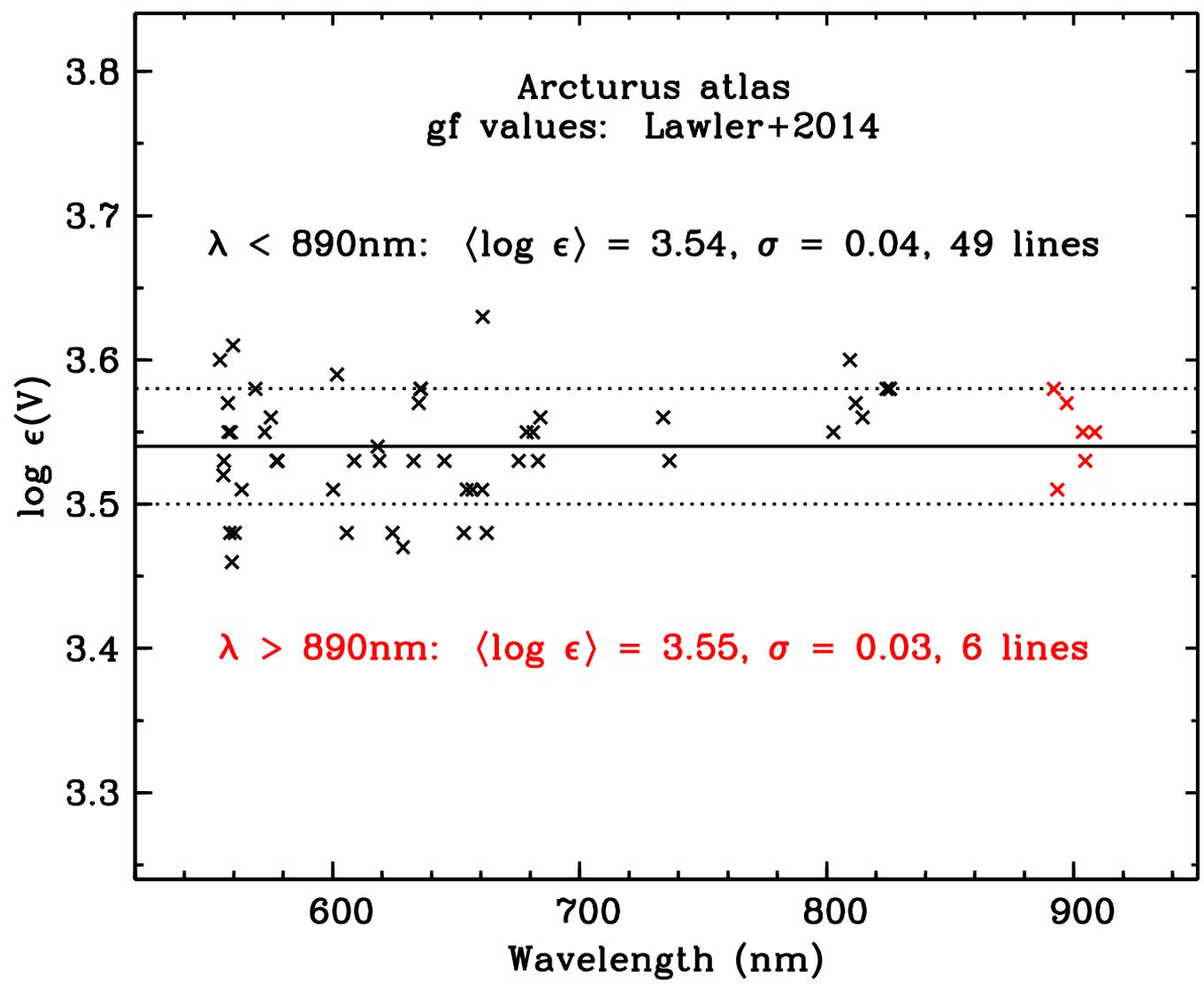

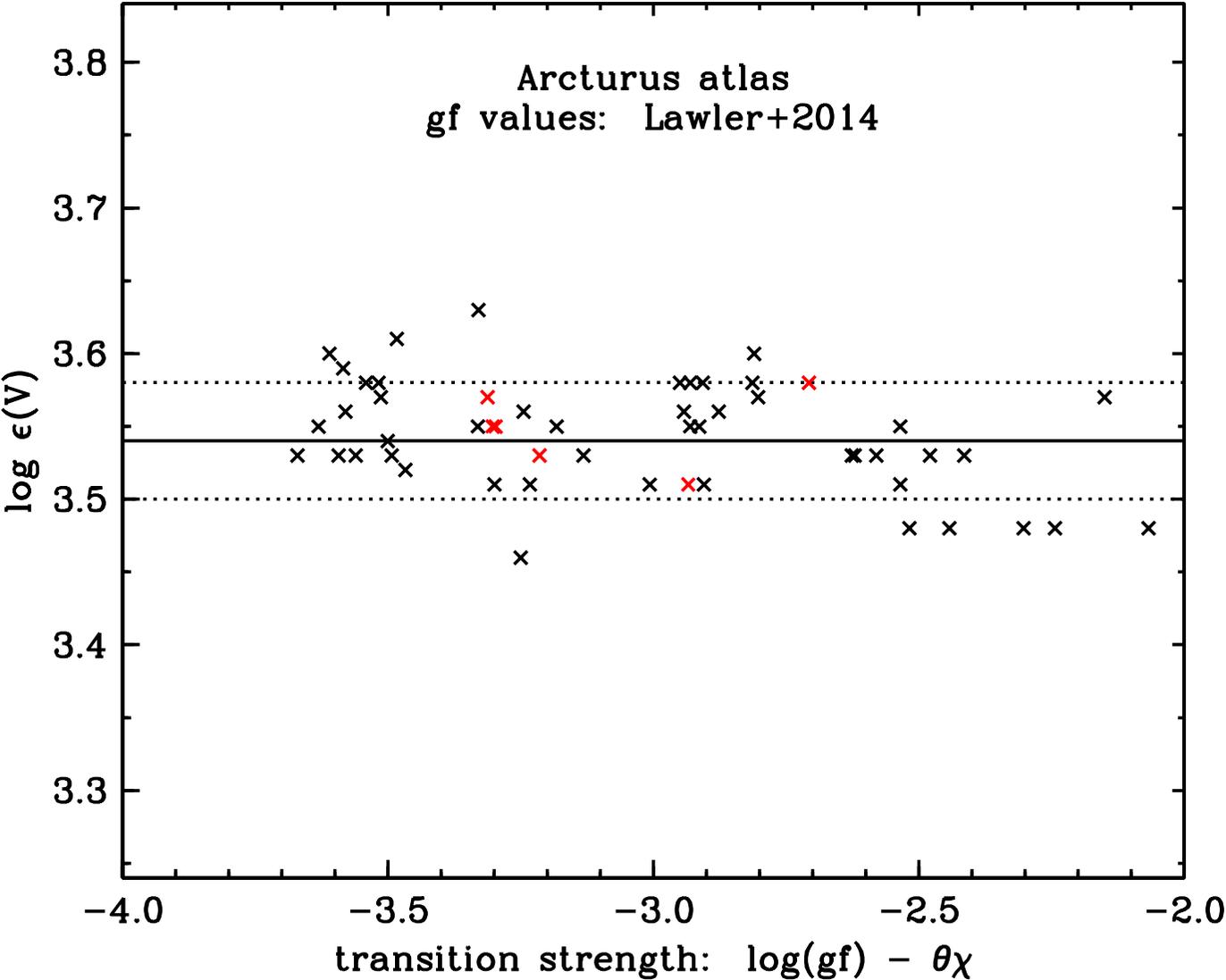

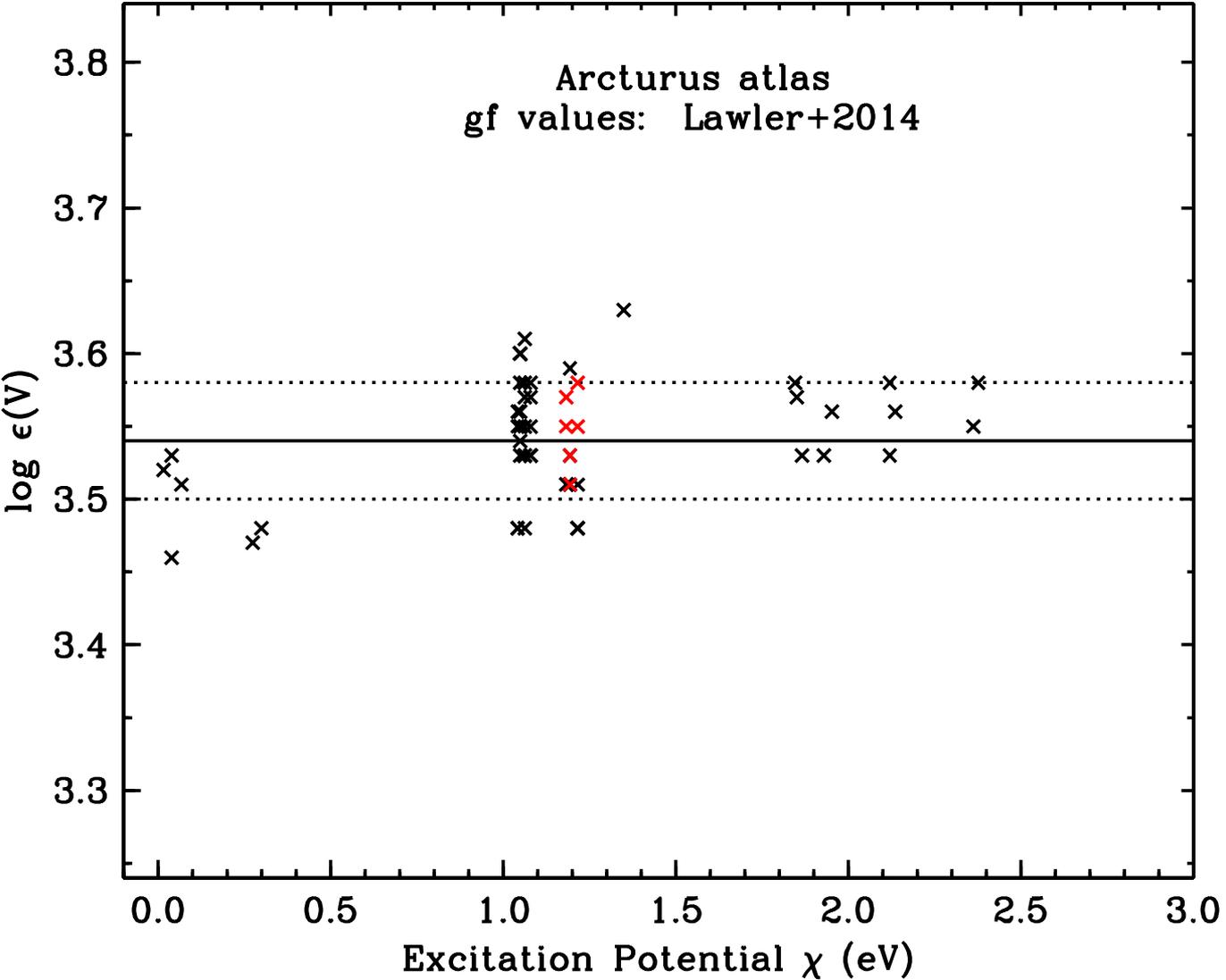

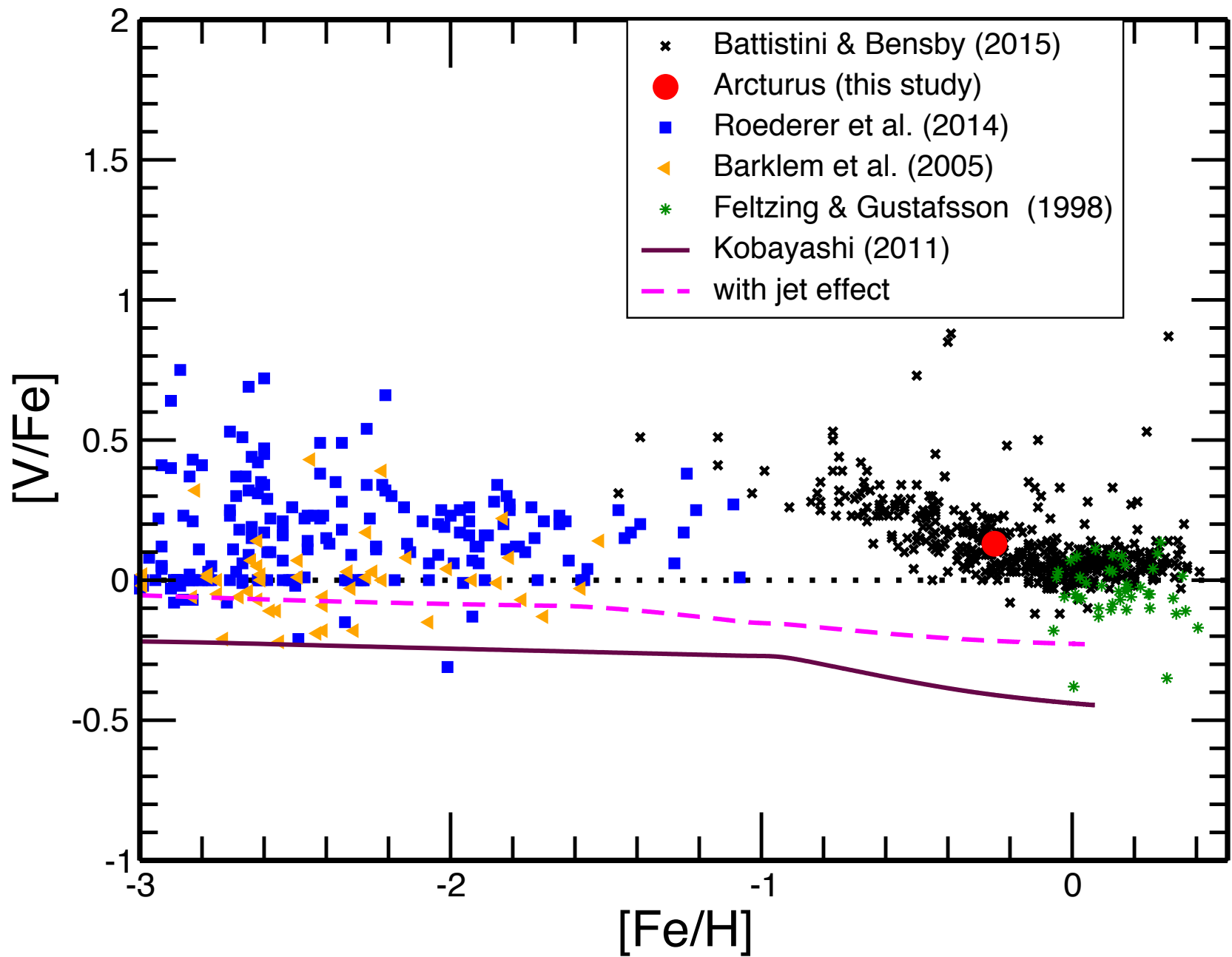

Table 3.  Hyperfine Structure (HFS) Line Component Patterns for 658 Transitions of $^{51}$V I.

| Transition Wavenumber[a] | Wavelength in air[a] | $F_{upp}$ | $F_{low}$ | Component Position | Component Position | Normalized Strength[b] |
|---|---|---|---|---|---|---|
| (cm$^{-1}$) | (nm) | | | (cm$^{-1}$) | (pm) | |
| 41620.972 | 240.1904 | 7 | 6 | 0.01867 | -0.1078 | 0.23437 |
| 41620.972 | 240.1904 | 6 | 6 | -0.04559 | 0.2631 | 0.03385 |
| 41620.972 | 240.1904 | 6 | 5 | 0.01876 | -0.1083 | 0.16927 |
| 41620.972 | 240.1904 | 5 | 6 | -0.10067 | 0.5810 | 0.00260 |
| 41620.972 | 240.1904 | 5 | 5 | -0.03632 | 0.2096 | 0.05320 |

Notes.  Table 3 is available in its entirety via the link to the machine-readable version online. These component patterns are computed from the best published HFS coefficients (see text) and energy levels from Thorne et al. (2011).

[a] Center-of-gravity wavenumbers and wavelengths are given, with component positions relative to those values.  Air wavelengths are computed using the standard index of air (Peck & Reeder 1972).

[b] Component strengths are normalized to 1.0 for each transition.